\documentclass[aps,prb,a4paper,reprint,superscriptaddress,showpacs]{revtex4-1}
\usepackage{graphicx,graphics,color,epsfig}
\usepackage{hyperref}
\usepackage{amsmath}
\usepackage{amsfonts}
\usepackage{amssymb}
\usepackage{appendix}

\begin{document}

\title{Spin filter and spin valve in ferromagnetic graphene}
\author{Yu Song}
\email{kwungyusung@gmail.com}
\affiliation{Institute of Electronic Engineering, China Academy of Engineering Physics,
Mianyang 621999, P.R. China}
\affiliation{Research Center for Microsystems and Terahertz, China Academy of Engineering Physics,
Mianyang 621999, P.R. China}

\author{Gang Dai}
\affiliation{Institute of Electronic Engineering, China Academy of Engineering Physics,
Mianyang 621999, P.R. China}
\affiliation{Research Center for Microsystems and Terahertz, China Academy of Engineering Physics,
Mianyang 621999, P.R. China}

\begin{abstract}
We propose and demonstrate that a EuO-induced and top-gated graphene ferromagnetic junction
can be simultaneously operated as a spin filter as well as a spin valve.
We attribute such a remarkable
result to a coexistence of a half-metal band and a common energy gap for opposite
spins in ferromagnetic graphene. We show that, both the spin filter and
the spin valve can be effectively controlled by a back gate voltage,
and they survive for practical metal contacts and finite temperature. Specifically, larger single
spin currents and on-state currents can be reached with contacts with
work functions similar to graphene, and the spin filter can operate at
higher temperature than the spin valve.
\end{abstract}

\date{\today} 
\maketitle

Graphene is a very promising material for the development of spintronics,
owing to a very weak spin-orbit interaction\cite{interaction}
and a rather strong field effect\cite{field-effect} in it. 
Recently, it is theoretically predicted that,\cite{proximity,FPC}
depositing a ferromagnetic (FM) insulator such as EuO on graphene
can induce short-range exchange interaction that
provides the spin potential difference needed for the operation of spintronics
in \emph{bulk} graphene.
The deposition of EuO on graphene has been experimentally realized and
its proximity induced ferromagnetization has been confirmed.\cite{Integration}
Also, FM graphene field effect transistors (FETs) have been made.\cite{FET}

To explore spin filter and spin valve functions in proximity-induced
FM graphene, spin polarization and magnetoresistance (MR) have been extensively
investigated.
It was found that, a spin polarized current is present in a single FM junction,\cite{proximity}
and the polarization can be enhanced 
in more effective structures (e.g., a superlattice\cite{superlattice1,superlattice2}
and a domain wall\cite{domainwall})
and modulated with extra modulations (e.g., magnetic,\cite{withmagnetic}
electric,\cite{withelectric} and strain\cite{withstrain} fields).
On the other hand, a MR can be achieved in a
double FM barrier structure,\cite{proximity} in which the equivalent magnetization of the barriers
are arranged to be parallel or antiparallel to each other.
The value of the MR can also be enhanced by resonant tunneling in superlattices.\cite{superlattice2}
Besides, suppression of evanescent waves by a second gate is also proposed to decrease the
minimum conductance in a FM junction.\cite{withelectric}
However, near 100\% polarization or almost vanished total conductance (i.e., a rather huge MR)
is still hard to achieve, even for ideal electrode contacts and at zero temperature,
limiting the realization of a perfect spin filter or spin valve in bulk graphene.

It should be noted that, all these works adopt a Zeeman splitting model
describing the short-range exchange interaction (see,
the left column in Fig. 1(a)),
which is roughly estimated through the analogy with a EuO/Al interface
and only opposite energy shifts are present for opposite spins.

In this Letter, we propose and demonstrate simultaneously achieved spin filter and
spin valve in a EuO-induced and top-gated graphene junction (see, Fig. 1(b)),
using a band structure recently revealed by first-principles
calculations (see, the right column in Fig. 1(a)).\cite{FPC}
We find that,
not only half-metal bands but also a common gap are present in the band
structure.
We show that, when the gate voltage is properly tuned that the Fermi
energy falls into the conduction or valence half-metal band
(case I or II in Fig. 1(c)),
the junction can function as a perfect spin filter;
while that the Fermi energy sweeps through the common gap (case III in Fig. 1(c)),
the junction functions as a perfect spin valve.
We also find that, the operation windows of
the spin filter and spin valve can be effectively tuned with the Fermi energy,
and the performance of both can be improved with a longer junction length.
We also show that, both the spin filter and spin valve are robust to
metal electrodes contact and finite temperature.

\begin{figure}[b]
\centering
\includegraphics[width=\linewidth]{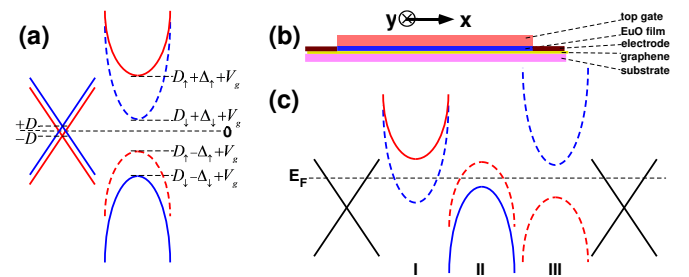}
\caption{(a) Comparison of the Zeeman splitting ($D\sim$5meV) and the FPC exposed band structures. 
(b) Schematic diagram (side view) of the proposed device.
(c) Schematic diagram of the mechanisms for the spin filter (case I
and II) and the spin valve (case III).}
\label{fig1}
\end{figure}

The proposed device is shown in Fig. 1(b).
A graphene strip with a dimension of $L\times W$ is placed in the $x$-$y$ plane,
on top of a SiO$_2$ substrate formed on the surface of a $n^{+}$Si back gate.
In a region of length $l(<L)$, a EuO layer is deposited on
the graphene strip,\cite{Integration}
and a top gate is further fabricated on top of the EuO layer.
An effective FM junction is formed
when the spin-resolved low-energy dispersion 
falls into the operation range of pristine graphene ($\pm1$eV around the charge neutral point).
This can be achieved by tuning the top gate voltage.
The junction is further contacted to metal electrodes
along the $y$-direction.
Due to difference in work functions, charge transfer occurs
between the metal and graphene.\cite{transfer}
This can be described by a potential barrier/well ($U$) in the electrode
region and a uniform Fermi energy through the device.\cite{contact}
The carrier concentration and the Fermi energy\cite{EF}
in the device can be tuned with the back gate through
the field effect.\cite{field-effect}
For non-suspended graphene, the transport
closed to the Dirac point is considerably affected by the elastic scattering
of substrate-induced Coulomb impurities.\cite{elastic,suspended}
To avoid this effect, we consider relatively large Fermi energies (i.e., 100-200 meV).
We also consider the case that the sample width
$W$ is much greater than the junction length $l$, 
so that the edge details of graphene are not important.\cite{WL}

The low-energy dispersion relation of FM graphene is found to be\cite{FPC}
\begin{equation}
E_s-D_s=\pm\sqrt{(\hbar v_s q)^2+(\Delta_s/2)^2},
\end{equation}
where not only Dirac points are shifted to $D_s=(-1.356+0.031 s)$ eV\cite{PC}
but also large Dirac gaps $\Delta_s=(58+9s)$ meV 
are opened, both depending on the spin index $s$
($=\pm1$ for up/down spin).
Besides, the Fermi velocities are re-normalized as $v_s=(1.4825-0.1455s)v_F$.\cite{note} 
We find that, the half-metal band exists between the two conduction band minimums (CBMs)
or the two valence band maximums (VBMs),
and the common gap is present between the higher VBM and the lower CBM.
In transport experiments on FM graphene FETs, it is observed that
the average charge neutral point shifts down with respect to pristine graphene,\cite{FET}
in qualitative agreement with the above result.

Noting that the dispersion is a combination of two independent 
dispersions of massive Dirac Fermions,
a combination of corresponding Hamiltonians\cite{review1} can be regarded as an effective Hamiltonian
of the FM graphene.
The result reads as
\begin{equation}
H_s=\sigma_0(D_s+V) +v_s\boldsymbol{\sigma}\cdot \mathbf{p}
+\xi\Delta_s\sigma_z,\label{eq1}
\end{equation}
where $\sigma_0$ is the identify matrix,
$V$ is the gate voltage,
$\boldsymbol{\sigma}=(\sigma
_{x},\sigma _{y})$ is the \emph{pseudospin} Pauli matrices
that originates from the hopping between different sublattices in graphene,\cite{pseudospin}
$\mathbf{p}=(\hbar k,\hbar q)$ is the momentum operator,\cite{spin-orbital}
and $\xi =\pm 1$ for valley $K$ and $K^\prime$.
For brevity, we express all quantities in dimensionless form by means of a
characteristic energy $E_0=10$ meV and corresponding length unit
$l_0=\hbar v_F/E_0=56.55$ nm.
The mean free math measured in SiO$_2$-supported graphene
at 100K is about 80 nm.\cite{suspended} It gets longer for suspended graphene\cite{suspended} and/or
at lower temperature ($T$).\cite{inelastic,elastic} In below calculations, we limit the junction length to
2$l_0$ for $T<$100K and 5$l_0$ for $T=0$, so that the transport
can be regarded as ballistic and only elastic scattering at the pristine-FM interfaces need to be considered.

The eigenstates in pristine and FM graphene ($j=e,m$)
can be exactly resolved by decoupling
the two-order differential equation $H_j\Phi_j=E_j\Phi_j$.
The result reads
$\Phi_j^\pm=[e^{\pm ik_jx},e^{\pm ik_jx}(\pm k_j+iq_j)/E_j]^T e^{iq_jy}/\sqrt{2}$,
where $E_e=E-U$, $E_m=(E-\widetilde{D}_s+\Delta_s)/v_s$ with
$\widetilde{D}_s=D_s+V$, $q_e=q_m=E_e\sin\alpha$ is the conserved transverse wave vector,
$k_e=\textmd{sign}(E_e)\sqrt{E_e^{2}-q_e^{2}}$, and
$k_m=\textmd{sign}(E_m)\sqrt{E_m E_m^\prime-q_m^{2}}$ with $E_m^\prime=(E-\widetilde{D}_s-\Delta_s)/v_s$.
By meaning of the matrix $U_j(x)=[\Phi_j^+(x),\Phi_j^-(x)]$,\cite{tuidao}
the envelope functions in the electrode and FM region can be written as
\begin{equation}
\Psi_j(x)=\frac{e^{iq_jy}}{\sqrt{2}}
\left(\begin{array}{cc}
e^{ik_jx} & e^{-ik_jx} \\
\frac{k_j+iq_j}{E_j}e^{ik_jx} & \frac{-k_j+iq_j}{E_j}e^{-ik_jx}%
\end{array}%
\right)
\left(
\begin{array}{cc}
u_j \\
v_j%
\end{array}%
\right),\label{eq2}
\end{equation}
where $u_j$ ($v_j$) is the complex coefficient for the right-
(left-) going propagation in region $j$.
For the incident region, $u_e=1$ and $v_e=r_s$, and for the transmission
region, $u_e=t_s$ and $v_e=0$.
Using the transfer matrix method,\cite{tmm}
the transfer matrix $M$ can be constructed as
$M=U_e^{-1}(l/2)U_m(l/2)U_m^{-1}(-l/2)U_e(-l/2)$ from
the boundary conditions $\Psi_e(-l/2)=\Psi_m(-l/2)$ and $\Psi_m(l/2)=\Psi_e(l/2)$.

The spin-resolved reflection and transmission coefficients,
$r_s=-M[[2,1]]M[[2,2]]^{-1}$ and $t_s=M[[2,2]]^{-1}$,
can be obtained as compact form
$r_s=(-B+C)\sin k_m l/(A\cos k_m l+B\sin k_m l)$
and $t_s=A e^{-ik_el}/(A\cos k_m l+B\sin k_m l)$,
where
$A=2i\lambda k_ek_m$, $B=E_e^2-2\lambda q_eq_m+\lambda^2E_mE_m^\prime$, 
$C=2k_e(k_e+iq_e-i\lambda q_m)$, and
$\lambda=E_e/E_m$ is the ratio between the effective electron energies
in the electrodes and FM regions.

Under the junction length condition we considered for ballistic transport,
inelastic (e-e and e-ph) scattering can be ignored,\cite{inelastic,phonon}
and the ballistic spin-resolved conductance at $T<$100K can be given
by the Landauer-B\"{u}ttiker formula\cite{LB}
\begin{equation}
G_s(E_{F},T)=G_{0}\int dE\frac{-df}{dE}\int_{-|E_{F}|}^{|E_{F}|}T_
s(E,q)\frac{dq}{2\pi /W},
\end{equation}%
where $T_s=|t_s|^2$ is the spin-resolved transmission probability,
$f(E,T)=[1+e^{(E-E_{F})/T}]^{-1}$ is the Fermi-Dirac distribution
function at Fermi energy $E_{F}$, and $%
G_{0}=2e^{2}/h$ is the quantum conductance (2 accounts for the valley
degeneracy). The zero-temperature conductance can be rewritten as
$G_s(E_{F},0)=MG_{0}\int_{-\pi /2}^{\pi /2}T_s(E_{F},\alpha )\cos \alpha
d\alpha$, where
\begin{equation}
M=(|E_{F}|/E_{0})(W/2\pi l_{0})\equiv M_EM_W
\end{equation}
is half of the number of the transverse modes.
The maximal channel conductance per spin is $2MG_{0}$.
The total conductance is $G=G_\uparrow+G_\downarrow$ and the spin
polarization is defined as $P=(G_\uparrow-G_\downarrow)/G$.

\begin{figure}[t]
\centering
\includegraphics[width=\linewidth]{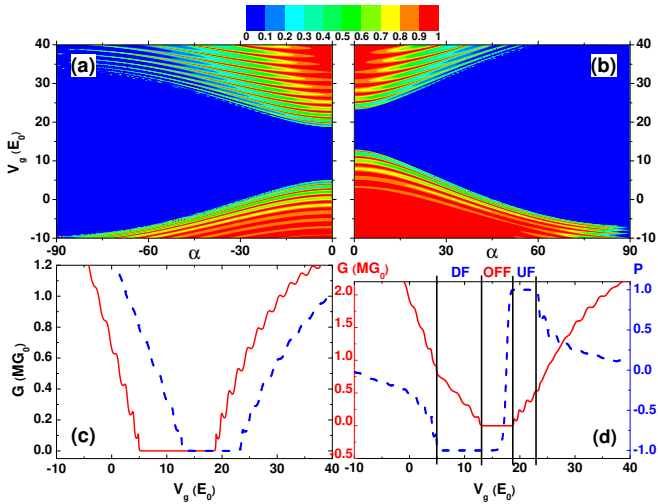} 
\caption{Contour plots of the (a) up and (b) down spin transmission probability
as a function of the gate voltage and incident angle. (c) Spin-resolved conductance,
(d) total conductance, and spin polarization
as a function of the gate voltage.
The parameters are $l=2$, $U=0$, $E_F=15$, and $T_P=0$.}
\label{fig2}
\end{figure}

Fig. 2 shows the calculation results for a junction of $l=2$
at a Fermi energy $E_F=15$.
The contact effect is ignored ($U=0$) and the temperature is zero.
It is observed that, each spin transmission displays two quasi-transparent
regions separated by a block one (Figs. 2a-b).
Meanwhile, a strong difference is present between up and down spins.
These results are clearly reflected in the spin conductance (Fig. 2(c))
and have a good correspondence to the band structure 
shown in Fig. 1(a).
Also observed are the stripes in the spin transmission and the
oscillations in the spin conductance,
which stem from the resonant tunneling of special $k_s[=(0,\pm1,\pm2,...)\pi/l]$.

Interestingly, between the two CBMs ($D_\downarrow+\Delta_\downarrow+V_g$
and $D_\uparrow+\Delta_\uparrow+V_g$),
the up spin conductance tends to zero and the down spin conductance remains
the order of the maximal channel conductance.
As a result, a perfect spin filter of spin down can be achieved,
which is clearly seen in the $P$-$E_F$ curve (see, the DF window in Fig. 2d).
The situation is just the opposite for Fermi energies falling between the two VBMs
($D_\downarrow-\Delta_\downarrow+V_g$ and $D_\uparrow-\Delta_\uparrow+V_g$),
where a perfect spin filter of spin up can be realized (UF window in Fig. 2d).
When the gate voltage is tuned that the Fermi energy enters
the common gap for both spins, i.e., $E_F\in(D_\uparrow-\Delta_\uparrow+V_g,D_\downarrow+\Delta_\downarrow+V_g)$,
the junction becomes `off' (see, the OFF window in Fig. 2d), as both spin are in their off-state (Fig. 2c).
For gate voltage out of this range, the junction becomes `on' because
at least one spin is in its on-state.
In total, the junction shows remarkable `on'-`filter'-`off'-`filter'-`on'
behaviour as the gate voltage changes within a certain range,
from which multifunction of spin filter and spin valve can be achieved.

A recent theoretical research by Beenakker's group\cite{valley} proposes and demonstrates that,
in a quantum point contact of zigzag graphene nanoribbons,
valley filter and valley valve can be \emph{respectively} achieved by local application of one or two
gate voltage to the point contact region.
For the latter case, the two gate voltages should be with opposite polarity.
Here in this work, we propose and demonstrate that, spin filter and spin valve can be
\emph{simultaneously} achieved in a graphene FM
junction by local application of a \emph{single} gate voltage.
Such a remarkable result stems from the fact that, not only a half-metal band
but also a common gap are present in FM graphene.
Besides, by using bulk graphene rather than graphene nanoribbons,
substantial (half of the maximum channel conductance) single spin current for the spin filter and on-state current
for the spin valve can be achieved (see, Fig. 2d),
which greatly benefits practical applications.

\begin{figure}[t]
\centering
\includegraphics[width=\linewidth]{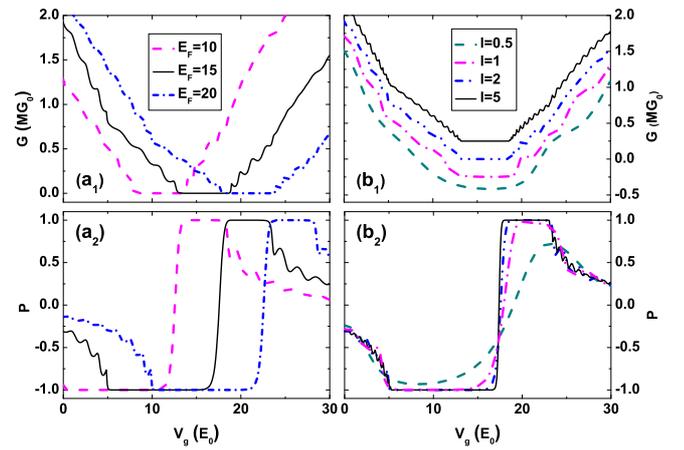}\\
\caption{Electrostatic control of the spin filter and the spin valve.
Total conductance and spin polarization as a function of the gate voltage
(a) for various Fermi energies at $l=2$ and
(b) for various junction lengths at $E_F=15$.
For clearness, the curves are shifted by 0.25 in (b$_1$).
$U=0$ and $T_P=0$ for all the cases.}\label{fig3}
\end{figure}

Electrostatic control is necessary for spintronics applications.
Fig. 3 shows the impact of the Fermi energy and junction length
on the performance of the spin filter and the spin valve.
As can be seen in Fig. 3(a),
the operation windows of both shift with the Fermi energy.
The single-spin current and on-state current are almost not affected.
As the junction length increases, the operation windows of the spin
filter (Fig. 3b$_2$) broaden to the band extremum limit and more oscillations appear
in the single spin current curve (Fig. 3b$_1$).
This is because, the blocked transport of up/down spin is further suppressed
and the quasi-transparent transport of down/up spin is further enhanced (Fig. 2c).
The performance of the spin valve is also advanced as the junction length increases (Fig. 3b$_1$).
Actually, too short a junction (e.g., $l$=0.5) can hardly work as an effective spin valve.
For spin valve, the blocked transport of both spin in the common gap (Fig. 2c)
are further suppressed as the length increases.

\begin{figure}[t]
\centering
\includegraphics[width=\linewidth]{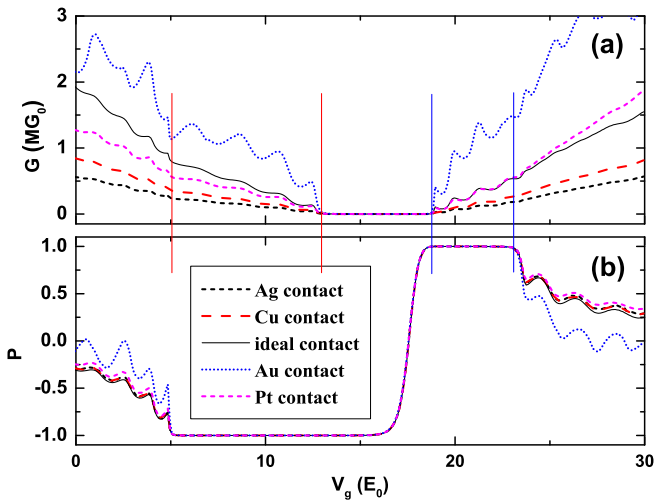}\\
\caption{
Electrode contact effect for the spin filter and the spin valve.
(a) Total conductance and (b) spin polarization as a function of the gate
voltage for various metal electrodes. $l=2$, $E_F=15$, and $T_P=0$.
}\label{fig4}
\end{figure}

In all the above calculations, the effect of electrodes
is ignored.
Such an ideal contact can be achieved by specific metals
with special distance to graphene (e.g., Au/Cu/Ag at 3.2/3.4/3.7${\AA}$).\cite{metal-contact}
Fig. 4 shows the calculated results for several familiar metal electrodes at their equilibrium
distances.
For Ag, Cu, Au, and Pt, $U/E_0=-32,-17,19,32$ respectively.\cite{metal-contact}
It is observed in Fig. 4(b) that, the two spin filter windows are almost
unaffected by the metal contact.
This is because the presence of $U$ and its value have no relations to
the spin degree of freedom.
It is not surprising that, the operation window of the spin valve is
not influenced by electrodes contact either (Fig. 4a).
Comparing with the ideal contact case, the single-spin current and on-state current
can even become bigger for electrodes with a small negative $U$ (e.g., Au).
These results suggest that, the graphene FM junction can simultaneously function as
spin filter and spin valve, even for a variety of metallic contacts.
It is also suggested that, to obtain a bigger single spin current or on-state current
electrode materials with smaller but similar work function as graphene should be adopted.

\begin{figure}[t]
\centering
\includegraphics[width=\linewidth]{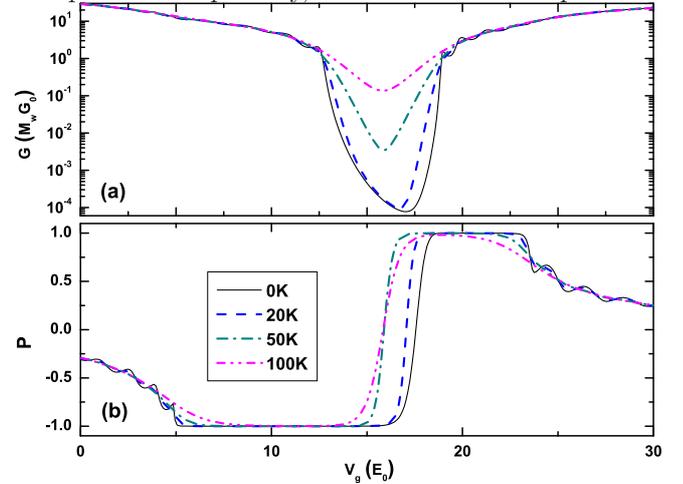}\\
\caption{Temperature effect for the spin filter and the spin valve.
(a) Temperature dependent total conductance and (b) spin polarization
as a function of the gate voltage at $E_F=15$.
For all the cases, the other junction parameters are $l=2$ and $U=0$.
Note $G$ is in unit of $M_W$ rather than
$M_EM_W$ adopted in Figs. 2-4 (see, Eq. 5).}\label{fig5}
\end{figure}

Finally, Figure 5 illustrates the effect of finite temperature.
It is observed in Fig. 5(b) that, the two spin filter windows narrow
down and shift to lower gate voltage as the temperature increases.
The up spin filter window even disappears at about 100K.
The main change of the single spin current is
the gradually smeared out of rich oscillations in the spectrum
at 0K (Fig. 5a).
For the spin valve, the `off' window narrows and disappears as the
temperature increases (Fig.5a).
The above behaviors can be understood from Eq.~(4).
Spin currents at a finite temperature $T$ are determined by the
zero-temperature spin currents in an
estimated energy range $\sim(E_{F}-5T_{P},E_{F}+5T_{P})$.
Generally, the spin currents (Fig. 2c) decrease (increase) with the temperature
in the quasi-transparent regions (blocked regions).
This leads to the increase of the total current in the common gap 
and the decrease of the polarization in the half-metal bands. 
The above results imply that, the proposed device can operate simultaneously as
a spin filter and a spin valve even at relatively high temperature.
Interestingly, the spin filter can survive for higher temperature than
the spin valve, because the former relies on the blocked transport of one spin
only, whereas the latter requires the blocked transport of both spins.

In summary, we have proposed and demonstrated that
simultaneous spin filter and spin valve can be achieved in
a EuO-induced and top-gated graphene junction,
provided a proper gate voltage is applied.
We have shown that, such a remarkable result arises because not only
a half-metal band but also a common gap are present in FM graphene,
which implies a great potential of the proposed junction for spintronics applications.
We have found that, the operation windows of the spin filter and the spin valve
can be effectively controlled by the Fermi energy, and
both the spin filter and the spin valve survive
for practical metal contacts and relatively high temperature.
Specifically, a larger single spin current and on-state current can be obtained
under contact metals with smaller but similar work function as graphene,
and the spin filter can survive for a higher temperature than
the spin valve.

YS acknowledges the support from NSFC under Grant No. 11404300.


\begin{references}
\bibitem{note}
Note, the above parameters are obtained for the optimized lattice and
the optimized distance between EuO and graphene.
\end{references}
\end{document}